\pdfoutput=1
\pdfoutput=1
\pdfoutput=1
\pdfoutput=1
\pdfoutput=1

\documentclass[fleqn]
{article}

\usepackage[T1]{fontenc}
\usepackage{amsfonts,amssymb,amsmath} % for math symbols.
\usepackage{graphics,graphicx,epsfig}
\usepackage{bm}% bold math
\usepackage[all]{xy}
\usepackage{color}
\usepackage{makeidx}
\makeindex
\usepackage{xr-hyper}
\usepackage[unicode,pagebackref]{hyperref}
\hypersetup{pdftex}

\usepackage{makeidx}
\usepackage{fancyhdr}
\fancypagestyle{plain}{%
    \fancyhead{}                      % ----на обычных страницах без колонтитулов

}
\newcommand{\beq}{\begin{equation}}
\newcommand{\eeq}{\end{equation}}
\newcommand{\beqa}{\begin{eqnarray}}
\newcommand{\eeqa}{\end{eqnarray}}

\date{\emph{North -Western Institute of Printing \\
of St-Petersburg State University  of Technology and Design,\\
Russia}
}

\title{Color to Gray and Back transformation for distributing  color digital images}

\author{V.N. Gorbachev, E.M. Kaynarova\footnote{E-mail:  helenkainarova@gmail.com},  I.K. Metelev,   E.S. Yakovleva \footnote{E-mail: 2305lena@mail.ru}.
}
\begin{document}

\maketitle

\begin{abstract}
 The Color to Gray and Back  transformation  watermarking with a  secrete key is considered. Color is embedded into the bit planes of  the luminosity component of the YUV  color space  with the help of  a block  algorithm that allows using not only the  least significant bits.  An application of  the problem of distributing   color digital images from a data base among legitimate users is discussed. The proposed protocol  can  protect original images  from  unauthorized copying.

\end{abstract}

%% The ``\keywordlist'' command prints out the keywords.

%\keywordlist

\section{Introduction}

Digital watermarking is one of the popular techniques  of  copyrighting  digital content.  However, this technique has a high potential  and  allows to solve some other problems particularly  the  Color to Gray and Back transformation.  The main idea of this transformation  is
to hide the color into a grayscale version of a color  image and then,
 (e.g.  after printing)  to extract the hidden  data  to retrieve  the original. Indeed, in practice this transformation is irreversible.

  Some applications of  Color to Gray and Back transformation are discussed in literature.  It may be suitable if e.g.  a color printer is not available and the document needs to be printed anyway \cite{1}. In this case someone might want to recover the colors later on from a printed black-and-white hardcopy. Next application is  a Hardcopy Image Backup System \cite{2}  proposed for archival storage of  photos or any  analog hard copies.  While a photo is stored
   the color can be changed  or faded and the System creates documents with hidden information about the initial color.  Also a color-shielded image transmission can be achieved if  color is embedded in the  randomly selected position   \cite{3}.  The set of positions is a secrete key  and only a legitimate user  which has the key  can recover the accurate color from a transmitted gray image.

The considered Color to Gray and Back transformation has the following  particular features.  Someone can get a grayscale  version of a color image  from its  representation,  e.g.  in  YCbCr  color space. The desired grayscale is   Y luminosity,  as the National Television Standards Committee recommends. The usual representation has 24 bit for color pixel and 8 bit for gray pixel.  Thus the transformation from color to gray   formally  is not reversible because entropy is not preserved.
In fact, a reversible transformation can be achieved  by watermarking  a color  into a grayscale image.  Formally the retrieved color image  can not be  equal to its original.   However  watermarking  provides two images that are undistinguishable  by  human visual system  and mapping  can be  viewed  as reversible.

 To embed  a color into  a grayscale image numerous techniques have been proposed, e.g.   DWT (Discrete Wavelet  Transform)  \cite{4}, consistent gradient field methods \cite{4a}, a virtual trit memory model \cite{4b}  and many  others.  Also information about  a color can be watermarked  in a  halftone of the printed  image  using  halftoning algorithms.  For example, a clustered- dot halftining may be applied  to carry information by shifting of the halftone clusters \cite{5}.

 In this paper  we consider digital color watermarking  with a secrete key.  In accordance with the Kerckhoffs's principle  the security of a system should reside in the chosen key.  At the same time the secret key allows to send   images to a legitimate user. The key can be chosen as a set of random positions in which  color is embedded  \cite{3}.  In contrast, we consider the key as  a random matrix that is used together with  our block coding  algorithm \cite{6} to hide color information.
 This algorithm  preserves the brightness of the block in which  a message bit is encoded. It results in enhancement  of the embedded data because  not only the least significant bits can be used.

  As a  practical application  of our approach  we consider  distributing  of digital color images from a data base. The problem is that any user could get the interested image before  buying. It seems to be  an important problem  because   many applications need  the hight quality photos  or color images (particulary images with MOS  that are needed  in image processing).  A solution can be found with the help of visible removable watermarking \cite{6.1}.  The idea is that  a visible watermark is embedded into an image and the interested user can remove it to retrieval unmarked image original  by  a secrete key that is available for an additional fee \cite{6.2}.  This is an analog of   a free-trial  version, when the interested user can stop running the trial period of time  by removing the watermark \cite{6.3}.
 In contrast  to the mentioned above idea  our  approach  uses an invisible watermark that is presented as a  hidden color.
 The proposed solution is based on the trade-off and exploits the  idea of $\beta$- version.  In fact,   a  grayscale version with a hidden color is available for the interested user  instead of  the original  color image .  We present a protocol  for retrieving  the original from its grayscale  $\beta$- version.
\\
The paper is organized as follows: first  we introduce the embedding algorithm, then an experiment and the protocol for   image  distribution are presented.

\section{Color Embedding}

The Color to Gray and Back transformation is an original example that shows how  digital watermarking technique works for  image processing applications.  We briefly discuss it  following  Queiroz et al \cite{4}.
\\
\\
The solution of  Queiroz  is based on DWT.  A RGB image is transformed into the YCbCr color space, where  Y  is a  luminosity and two chrominance components Cb and Cr keep  the color information. One level DWT of luminosity  DWT(Y) has four blocks,   the size of each block  is  a quarter of  Y. Two blocks  are replaced by  two quarters of coefficients from Cr and Cr and a new luminosity Y' appears.  Then an inverse  transform  of Y' results in  a graysacle version  Yc=IDWT(Y') with  color embedded. To  recover the  color we  need to revers the mentioned above  steps.   According to this approach a quater of all color information was exploited. That means that the initial and retrieved color images are not equal  however they are visually  undistinguishable.
\\
\\
In our model  an RGB image  is transformed into YUV color space, where Y is a luminosity and $X=U,V$ are two chrominance components.  Using a  secret key $K$ we create a message
  \begin{eqnarray}
  \label{001}
  % \nonumber to remove numbering (before each equation)
    M=(K+X)\   mod(255),
  \end{eqnarray}
where $K$ is a random matrix.   Then $M$ is embedded into luminance Y, that represents  a coverwork.
\\
\\
To embed $M$ we use our algorithm \cite{6} that keeps  the brightness of an image  as far as  it is possible. Thus more  information  can be embedded  without introducing noticeable distortions.  The algorithm uses the $Y$  bit planes  and works as follows.  Each bit plane  $Y_{V}$ has weight  $2^{V-1}$,  $V=1,\dots, 8$ and  it  is divided into a set of non-overlapping blocks  $Y_{Va}=\{y[m,n]\}$  of $h\times h$  size.  Let us introduce a parity bit of the block  side diagonal
\begin{eqnarray*}
% \nonumber to remove numbering (before each equation)
  d_{a}=\bigoplus_{x\in Y_{Va}} y[x, h-x].
\end{eqnarray*}
A bit   of the message $m$ is encoded by  the block $Y_{Va}$  as follows:
\begin{eqnarray*}
% \nonumber to remove numbering (before each equation)
  E: \ Y_{Va}\to S_{Va}=\left\{
               \begin{array}{ll}
                 Y_{Va}, ~~~~~if~ d_{a}\oplus m=0,& \\
                 ZY_{Va}, ~~~if~ d_{a}\oplus m=1. &
               \end{array}
             \right\},
\end{eqnarray*}
where the operator $Z$  either modifies a bit of the side diagonal  or  finds a block $S_{Va}$
whose brightness is equal or closest to the  brightness of  $Y_{Va}$.  So brightness of luminance Y is preserved as far as it is possible.
\\
The next example illustrates how operator  $Z$  works in  case of  $h=2$.
  \\
     \begin{eqnarray*}
    %\nonumber to remove numbering (before each equation)
    &&M=\begin{bmatrix}
        \mathbf{ 1} & \mathbf{1} \\
          \mathbf{0} & \mathbf{1} \\
       \end{bmatrix}
       \to Y_{Va}=\begin{bmatrix}
                            0 & 1~\vline   ~1 & 1 \\
                            0 & 1 ~\vline  ~1 & 1 \\
                            \hline
                            0 & 1~\vline  ~0 & 1 \\
                            0 & 1 ~\vline ~ 1 & 0 \\
                         \end{bmatrix};
 %\end{eqnarray*}
%\begin{eqnarray*}
%&&
S_{Va}=\begin{bmatrix}
                            0 & \mathbf{1}~\vline   ~1 & \mathbf{0} \\
                           \mathbf{ 0} & 1 ~\vline  ~\mathbf{1} & 1 \\

                            \hline
                            0 &  \mathbf{1}~\vline  ~1 &  \mathbf{1} \\
                                \mathbf{1} & 0 ~\vline ~  \mathbf{0} & 0 \\
                         \end{bmatrix}.
        \end{eqnarray*}
        \\
Here  a message $M$ is a  $2\times 2$ matrix.  Each bit of $M$ is embedded into one of four blocks of
 $Y_{Va}$.  The stegoimage $S_{Va}$  consists of four blocks and only one of them, the top right block, has its brightness that is change to 1.
\\
A detection algorithm is blind. A  hidden message is extracted from the bit plane of the stegoimage
$S_{Va}$ by calculating the parity bit $d_{a}=m$.    Then  new chrominance  components $X'=U',V'$ are retrieved  using the secrete key
\begin{eqnarray*}
% \nonumber to remove numbering (before each equation)
  X'=(M-K+255)\   mod(255).
\end{eqnarray*}
Indeed,  a bit plane up to $V=5,6$ can be used without introducing  any visual changes.

\section{Experiment}

\begin{figure}[!h]
  \begin{center}
     \includegraphics[width=15cm,height=15cm]{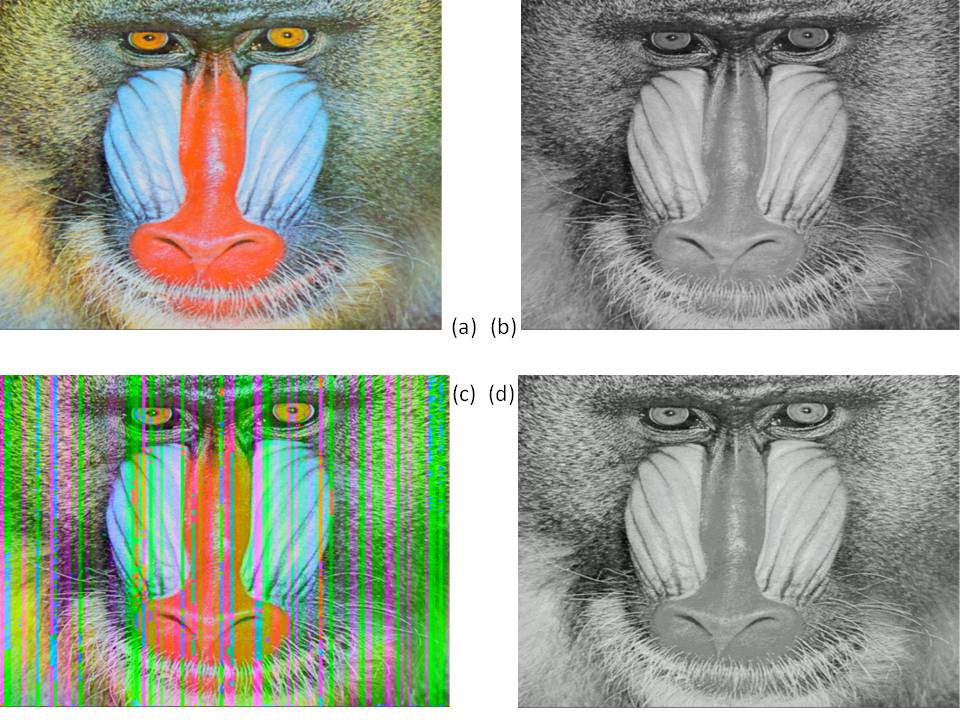}
  \end{center}
  \caption{Color embedding.  (a) Color original; (b) coverwork image, luminosity component  $Y$ of  $YUV$ color space; (d) luminosity $Y$  with hidden color; (c) the retrieved original with slightly corrupted secrete key.}\label{Paper}
\end{figure}

The initial RGB image and its  luminosity Y  of  the YUV  color space are presented at  Figure
\ref{Paper}.
\\
Amount of bits  $H_{Y}$ available for embedding  depends on the number of the bit planes $T$, the number of total pixels $n$ and the size of the block $h$ as $H_{Y}=nT/h^{2}$. It is obvious, that all color information can't be hidden into Y.
Then we decimate chrominance components  $U,V$ by averaging over the $u\times u$ environment.  For $u=4$ amount of color  information  $H_{C}$ is $1/8$ from $Y$ or $H_{C}=(1/8)nk$,  where  $k$  is a bit depth of the channel  $Y,U, V$ and  usually  $k=8$.  If $h=2$ we find  $H_{Y}=(T/4)H_{C}$. It means that all color information $H_{C}$ can be embedded if $T=4$,  in another words  it needs all Y  bit planes be  up to $V=4$. Figure \ref{Paper}-(d) ishows the luminosity component with  color embedded into  four bit planes $V=1-4$.

\begin{figure}[!h]
  \begin{center}
     \includegraphics[width=15cm,height=15cm]{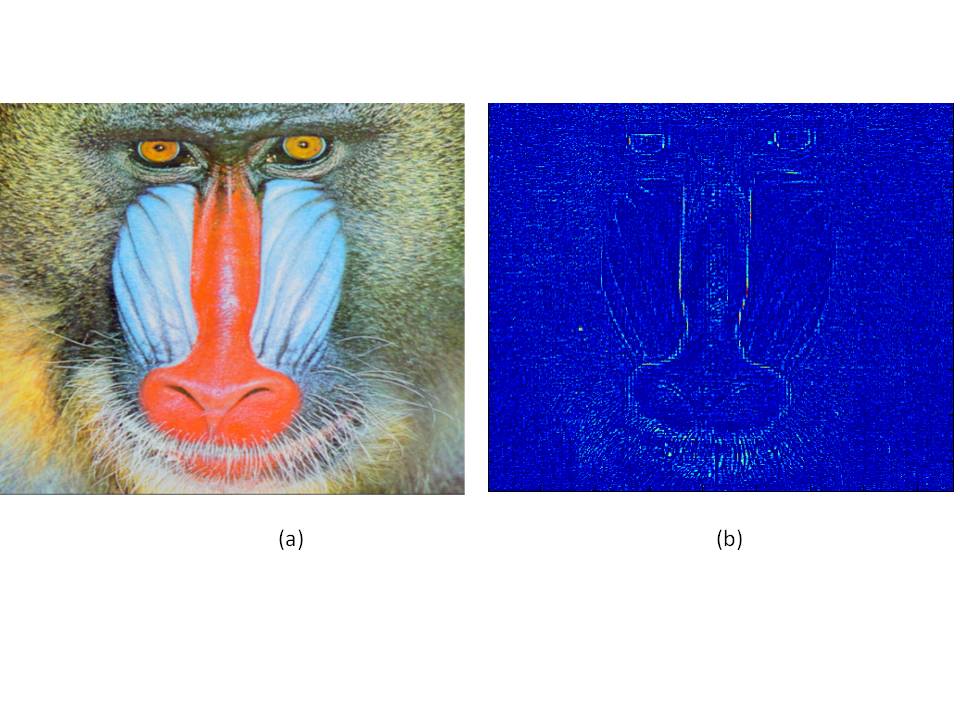}
  \end{center}
  \caption{  (а)  Retrieved color image; (b) difference in  blue channels between the original and the retrieved images.}\label{cPaper}
\end{figure}

\begin{figure}[!h]
  \begin{center}
     \includegraphics[width=15cm,height=17cm]{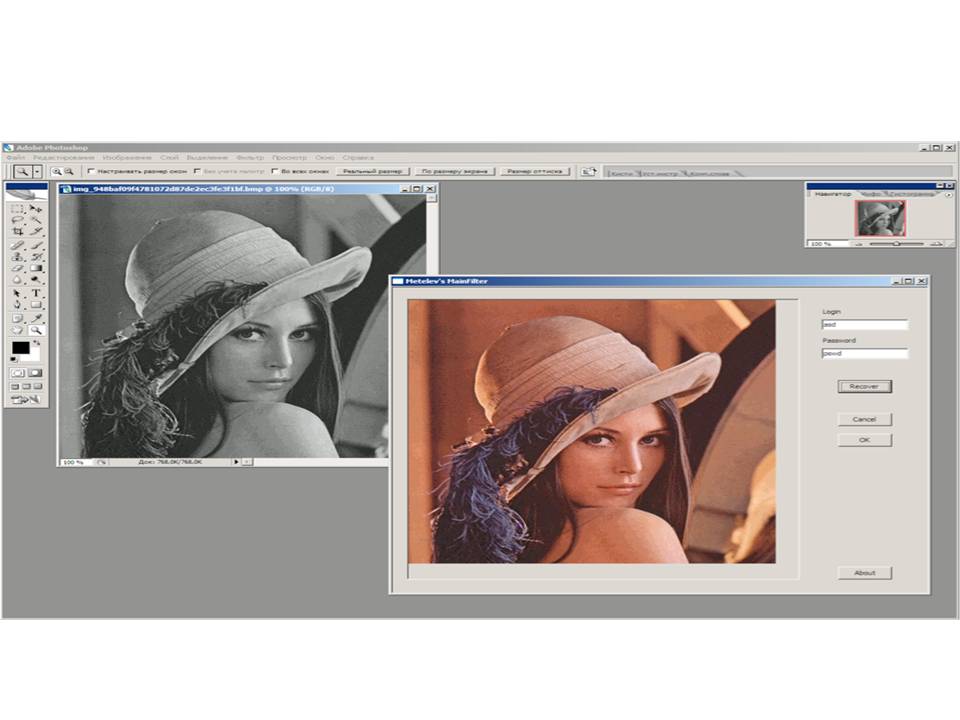}
  \end{center}
  \caption{Adobe Photoshop Interface.}\label{ccPaper}
\end{figure}

To retrieve the initial RGB image we need the secret key $K$. If $K$ is unknown, it can't be found in practice. The reason is that
searching of  $2^{nk}$ matrix is a hard problem with  the  non polynomial (NP) computational complexity.   Fig. \ref{Paper}-(c) illustrates an example when  color  is retrieved without  the secrete key so  we take a corrupted key, that has some bits changed.  A large number of distortions can be found. If  the secrete key is known the retrieved color image seems to be  visually undistinguishable from its original, as shows Figure
\ref{cPaper}-(a).  However there are  invisible differences,  they are shown at  Figure \ref{cPaper}-(b)  as a blue component.
\\
\\
To illustrate the proposed approach  a user interface associated with $Adobe Photoshop$ is presented at Figure
\ref{ccPaper}.

\section{Protocol of color images distributing}

Let us introduce our approach with a  secrete key to  the problem of   digital images distributing  among  legitimate users. It may address to hight quality photos or data base of unique images distributing.
\\
The problem  is as follows. \emph{ Alice has a digital color image and she wishes to sell it.  Bob wishes to by it but before this  he wants to get it to know; however Alice distrusts Bob.
} The  solution can  found using the following  protocol.
 Instead of transmitting the original color  image $A$ Alice sends to Bob its grayscale version $B$ with hidden color.  After Bob informs Alice about  his decision Alice sends him the secrete key and Bob retrieves the desired color image.
\\
As a result Alice keeps her copyright because she uses the secrete key and Bob can get the interested image to know it, but this will be  a grayscale version.  Let us consider another solution.  Alice has two images, an original color image $A$ and its grayscale version $B'$ for free distributing.  In this case both images $A$ and $B'$ have to be transmitted through the channel to Bob.  Then any criminal Eve can get color image $A$  attacking  the channel. According to our model  it is impossible. The color image is not  transmitted trough the channel.  As for sending the secrete key to Bob, this is a well known problem of key distribution and Alice can use one of the standard protocols of cryptography.
\\
\\
We are grateful to  Kharinov M. V.  for hopeful discussion.

\section{Conclusions}
Steganographic techniques allows us to solve the problems   of  information   protection and many others.

 During image processing there are numerous   transformations of color image into gray.  In practice  these transformations are irreversible. However reversibility can be examined from the point of  view  of human vision.  So, a transformation  can be considered as reversible if original and its retrieval version are undistinguishable by human eyes.  And we can  find that Color to Gray and Back transformation is reversible.
 \\
Applications of  Color to Gray and Back transformation
refer to printing when color is retrieved from a halftone hard copy. Actually  printing  makes
the color detection very hard so we consider an example without printing.  Our example refers to the problem   of color  digital images or  photos from a data base  distribution.  The goal is  that  any user can  get the interested photo to know before  buying.  The proposed solution is based on  the idea of $\beta$- version.  Instead of the original image, its grayscale version with hidden color  is  available,   then the legitimate user could retrieve the interested color photo using the secrete key.

\newpage
%%%%%%%%%%%%%%%%%%%%%%%%%%%%
%%%%%%%%%%%%%%%%%%%%%%%%%%%%%

\end{document}